\documentclass[12pt]{iopart}
\usepackage{epsf}  
\begin{document}
\newcommand{\erti}{Er$_2$Ti$_2$O$_7$}
\newcommand{\tbti}{Tb$_2$Ti$_2$O$_7$}
\newcommand{\er}{Er$^{3+}$}
\newcommand{\mub}{$\mu_B$}
\title{Magnetisation process in \erti\ and \tbti\ at very low temperature}

\author{P. Bonville\ddag\footnote[3]{To whom correspondence should be addressed (pierre.bonville@cea.fr)}, S. Petit\dag, I. Mirebeau\dag, J. Robert\dag, E. Lhotel\S\S\ and C. Paulsen\S\S 
}

\address{\dag\ CEA, Centre de Saclay, Laboratoire L\'eon Brillouin, 91191 
Gif-sur-Yvette, France}

\address{\ddag\ CEA, Centre de Saclay, IRAMIS/Service de Physique de l'Etat
Condens\'e, 91191 Gif-sur-Yvette, France}

\address{\S\S\ Institut N\'eel, CNRS \& Universit\'e Joseph Fourier, 38042 Grenoble, France}

\begin{abstract}
We present a model which accounts for the high field magnetisation at very low temperature in two pyrochlore frustrated systems, \erti\ and \tbti. The two compounds present very different ground states: \erti, which has a planar crystal field anisotropy, is an antiferromagnet with $T_{\rm N}$=1.2\,K, whereas \tbti\ is expected to have Ising character and shows no magnetic ordering down to 0.05\,K, being thus labelled a ``spin liquid''. Our model is a mean field self-consistent calculation involving the 4 rare earth sites of a tetrahedron, the building unit of the pyrochlore lattice. It includes the full crystal field hamiltonian, the infinite range dipolar interaction and anisotropic nearest neighbour exchange described by a 3-component tensor. For \erti, we discuss the equivalence of our treatment of the exchange tensor, taken to be diagonal in a frame linked to a rare earth - rare earth bond, with the pseudo-spin hamiltonian recently developped for Kramers doublets in a pyrochlore lattice. In \tbti, an essential ingredient of our model is a symmetry breaking developping at very low temperature. We compare its prediction for the isothermal magnetisation with that of ``the quantum spin ice'' model.

\end{abstract}


\submitto{\JPCM}

\section{Introduction}
The pyrochlore titanates, with formula R$_2$Ti$_2$O$_7$ where R is a rare earth, have been the subject of intense studies for a decade \cite{gardging}. The pyrochlore lattice, made of tetrahedra joined by their vertices, leads indeed to a frustration of the exchange interaction in some specific situations. Owing to different crystal field properties, to the Kramers or non-Kramers character, to the relative importance of the dipolar interaction with respect to exchange, these compounds display a great variety of low temperature behaviours, the best known being the ``spin-ice'' ground state occurring in Ho$_2$Ti$_2$O$_7$ and Dy$_2$Ti$_2$O$7$ \cite{harris,ramirez}. In the last years, there has been a break-through towards a full understanding of the low temperature properties of the Tb, Er and Yb members of the series, in particular with the recognition that the anisotropy of the nearest neighbour exchange interaction plays an essential role \cite{cao09,thompson,onoda}. We are interested here in the very low temperature field variation of the magnetisation in \erti\ and \tbti, and we will show that much of the experimental data available in the two compounds can be accounted for, in each case, using a unique anisotropic exchange tensor.

\erti\ is an antiferromagnet with $T_{\rm N}$=1.2\,K \cite{blote} and it presents an easy magnetic plane, perpendicular to the local $<111>$ ternary axis, arising from the crystal field ground doublet of the Kramers Er$^{3+}$ ion \cite{cao09,siddh}. The magnetic structure has been determined \cite{champion,poole} and magnetic order was suggested to arise from the ``order by disorder'' mechanism \cite{villain}, which was recently put on a more robust ground \cite{zhito,savary}. The phase diagram and the evolution of the magnetic structure upon application of a magnetic field were also determined \cite{ruff,cao10} and interpreted in terms of anisotropic exchange, using a simple model \cite{cao10}. A symmetry constrained 4-component anisotropic exchange tensor was eventually introduced \cite{savaryx} and derived from fitting of the spin wave dispersion laws \cite{savary}. In these latter works, the hamiltonian is written in terms of effective S=1/2 pseudo-spins which represent the ground doublet alone. This approach is adequate to describe the physics of \erti\ at low temperature (the two first excited doublets have an energy of 73\,K and 85\,K above the ground state \cite{champion}) and in zero or moderate magnetic fields (a few T, of the order of the critical field $H_c \simeq$ 1.7\,T, see below), where mixing with the excited doublet is negligible. The pseudo-spin hamiltonian cannot however capture the high field magnetisation since mixing plays herein an important role. In the first part of this work, we start by showing the equivalence of the 4-component pseudo-spin exchange tensor and of the exchange tensor used in the present work, which is diagonal in a R-R bond frame. Then we show that the zero-field antiferromagnetic (AF) ground state in \erti\ and the single crystal magnetisation curves in the AF phase are correctly described by mean field theory, considering the full crystal field interaction, involving thus the total rare earth momentum {\bf J}, and anisotropic exchange. We show that the same exchange tensor also allows to reproduce the neutron diffuse scattering in the paramagnetic phase.

In \tbti, a quite different picture is relevant. The ground state presents no long range magnetic ordering down to very low temperature \cite{gardner}. The crystal field level scheme of the non-Kramers Tb$^{3+}$ ion has two ground doublets separated by a small gap $\Delta \simeq$1.4\,meV \cite{gingras,mirebeau}, each having Ising anisotropy along the $<111>$ ternary axis. The exchange in \tbti\ is of AF type, and thus the lack of magnetic ordering is puzzling since antiferromagnetic exchange is not frustrated with Ising-like spins on the pyrochlore lattice. A model involving quantum fluctuations between the two ground doublets \cite{molavian} was developed to account for the ``spin-liquid'' behaviour in \tbti. This model predicts that the very low temperature magnetisation curve should present a plateau (for a field along [111]) around 0.05\,T \cite{molavianging}, akin to the ``2/3 plateau'' observed in spin-ices \cite{matsu}. However, two recent works \cite{dunsiger,lhotel} did not confirm this prediction. An alternative model for \tbti\ was proposed by us \cite{bonv11,bonville}, based on the occurence of a symmetry breaking at low temperature, caused for instance by a Jahn-Teller distortion of the rare earth site or by quadrupolar ordering. In a certain region in the space of exchange integrals, this approach yields a phase with no long range magnetic ordering. In the second part of this work, we show that the very low temperature magnetisation curves in a single crystal \cite{dunsiger,lhotel} can be reproduced, to a good approximation, in the frame of our model, using the anisotropic exchange tensor derived in Ref.\cite{bonville}.

\section{The exchange tensor and the magnetisation in \erti}

\subsection{The anisotropic exchange tensor} \label{anis}

Various conventions have been used to define the anisotropic exchange tensor in the pyrochlores. Since exchange is a two-ion interaction, a natural choice is to consider the vector linking two rare earth ions (the R-R bond) as principal axis for the exchange tensor \cite{malkin}. We make this choice in the present work, i.e. we take a 3-component exchange tensor $\tilde {\cal J}$ which is diagonal in a ``bond frame'' ({\bf a},{\bf b},{\bf c})  where {\bf c} is the R-R bond axis (see the Appendix for a definition of the bond frames). One can also add an antisymmetric Dzyaloshinsky-Moriya exchange term, with parameter $J_4$. In this frame, the exchange interaction between ions $i$ and $j$ writes:
\begin{equation}
{\cal H}_{ex}^{ij} = - [ {\cal J}_a \ J^i_x . J^j_x + {\cal J}_b \ J^i_y . J^j_y + {\cal J}_c \ J^i_z . J^j_z + \sqrt{2}\ J_4\ ({\bf J}^i \times {\bf J}^j)_y],
\label{diagex}
\end{equation}
where the $J^i_\alpha$ are the components of the full angular momentum. Another choice is made in Refs.\cite{savary,savaryx}, where an S=1/2 pseudo-spin is considered to describe the ground doublet and the exchange hamiltonian is written in terms of the pseudo-spin components. In case of a Kramers doublet in the local frame with trigonal symmetry, the 3-component axial g-tensor writes $\tilde g$ = \{$g_\perp,g_\perp,g_z$\}, and the projection onto the ground doublet yields the relationship: $g_{\rm J}${\bf J} = $\tilde g$ {\bf S}, where $g_{\rm J}$ is the ionic Land\'e factor. 

In the Appendix, we demonstrate that these two formulations of the exchange hamiltonian are equivalent and we show how to derive the relationships between the \{${\cal J}_a,{\cal J}_b,{\cal J}_c,J_4$\} tensor used here and the exchange parameters \{$J_{zz},J_\pm,J_{\pm\pm},J_{z\pm}$\} entering the exchange hamiltonian in terms of pseudo-spins. They write as follows (for a Kramers doublet):
\begin{eqnarray} \label{trans}
J_{zz} & = & (\frac{g_z}{g_{\rm J}})^2\ \frac{-{\cal J}_a +2{\cal J}_c+4 J_4}{3} \nonumber \\
J_\pm  & = &(\frac{g_\perp}{g_{\rm J}})^2\  \frac{2{\cal J}_a-3{\cal J}_b-{\cal J}_c+4 J_4}{12} \\
J_{z\pm}  & = &-\frac{g_z g_\perp}{g_{\rm J}^2}\  \frac{{\cal J}_a+{\cal J}_c-J_4}{3\sqrt{2}} \nonumber \\
J_{\pm\pm}  & = &(\frac{g_\perp}{g_{\rm J}})^2\  \frac{-2{\cal J}_a-3{\cal J}_b+{\cal J}_c-4J_4}{12} \nonumber
\end{eqnarray} 
The reciprocal relations write:
\begin{eqnarray} \label{snart}
{\cal J}_a & = & \frac{1}{3}[-(\frac{g_{\rm J}}{g_z})^2\ J_{zz} +4(\frac{g_{\rm J}}{g_\perp})^2(J_\pm-J_{\pm\pm})- 4\sqrt{2} \frac{g_{\rm J}^2} {g_zg_\perp} J_{z\pm}] \nonumber \\
{\cal J}_b & = & -2(\frac{g_{\rm J}}{g_\perp})^2\ (J_\pm + J_{\pm\pm}) \\
{\cal J}_c & = & \frac{2}{3}[(\frac{g_{\rm J}}{g_z})^2\ J_{zz} -(\frac{g_{\rm J}}{g_\perp})^2\ (J_\pm-J_{\pm\pm})- 2\sqrt{2}\frac{g_{\rm J}^2}{g_z g_\perp}\ J_{z\pm}]
 \nonumber \\
J_4 & = & \frac{1}{3}[(\frac{g_{\rm J}}{g_z})^2\ J_{zz} +2(\frac{g_{\rm J}}{g_\perp})^2\ (J_\pm-J_{\pm\pm}) + \sqrt{2}\frac{g_{\rm J}^2}{g_z g_\perp}\ J_{z\pm}].
 \nonumber
\end{eqnarray}
We emphasize that the physical parameters here are ${\cal J}_a$, ${\cal J}_b$, ${\cal J}_c$, $J_4$ and the two components $g_z$ and $g_\perp$ of the g-tensor, the pseudo-spin constants being effective parameters.
 
\subsection{Isothermal magnetisation in the AF phase and diffuse scattering in the paramagnetic phase}

\begin{figure}
\epsfxsize=430pt
\center{
\epsfbox{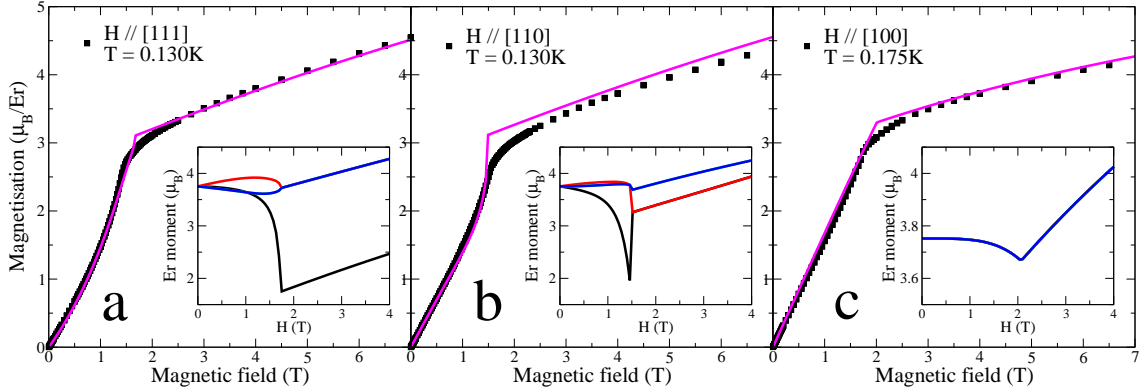}
\caption{\label{ertimh} Isothermal magnetisation curves in \erti\ close to 0.15\,K, with the magnetic field applied along [111] ({\bf a}), [110] ({\bf b}) and [100] ({\bf c}): the black dots are the experimental data and the solid line the calculated curve. Insets: calculated moduli of the Er moments for the 4 sites in a tetrahedron. For {\bf H} // [111], the black curve corresponds to the Er site with its ternary axis parallel to [111], the other colors to the 3 other sites. For {\bf H} // [110], the black and red curves correspond to the $\alpha$-sites, for which the local axis is at an angle of acos($\sqrt{\frac{2}{3}}$) $\simeq 35.3^\circ$ from the field, and the blue curve to the $\beta$-sites, for which the local axis is perpendicular to the field. For {\bf H} // [100], the 4 moduli have the same field variation. All the simulations were performed with the anisotropic exchange tensor ${\cal J}_a =0.030$\,K, ${\cal J}_b=-0.050$\,K, ${\cal J}_c=-0.105$\,K and $J_4$=0.}
}
\end{figure}

The magnetisation $vs$ field has been measured in the AF phase of an \erti\ single crystal, at 0.15\,K, using a SQUID magnetometer equipped with a dilution refrigerator developed at the Institut N\'eel-CNRS. The magnetic field was applied along the 3 symmetry directions [111], [110] and [100], and the data are shown as black dots in Fig.\ref{ertimh}. The demagnetisation factor was negligible for the 3 directions. For {\bf H} // [111] and [110] (Figs.\ref{ertimh} {\bf a} and {\bf b}), one observes a steep increase of the magnetisation with a slight upwards curvature up to a critical field $H_c \simeq$1.6\,T; for {\bf H} // [100] (Fig.\ref{ertimh} {\bf c}), the initial increase is linear and the critical field slightly higher: $H_c \simeq$1.8\,T. Above $H_c$, a much slower linear increase is observed up to the maximum field of 7\,T. We have computed the field variation of the magnetic structure in each case, using a self-consistent mean field calculation which involves the 4 Er sites of a tetrahedron, each being exchange coupled to its nearest neighbours, and taking into account the infinite range dipole-dipole interaction. The underlying asssumption of this type of calculation is that the magnetic structure under field has a {\bf k}=0 propagation vector. This is the case when the field is applied along the [110] direction \cite{ruff,cao10} and it is probably so for {\bf H} // [111] and [100], as can be concluded {\it a posteriori} from our study. The \er\ ion (J=15/2, $g_{\rm J}$=6/5) is described by its full angular momentum {\bf J} submitted to a trigonal crystal field interaction with parameters as in Ref.\cite{cao09}. The g-tensor of the ground doublet has components: $g_z$=2.6 and $g_\perp$=6.8 \cite{cao09}. 

Starting from the pseudo-spin exchange parameters derived for \erti\ in Ref.\cite{savary}, relations (\ref{snart}) allow us to obtain the exchange integrals ${\cal J}_a$, ${\cal J}_b$, ${\cal J}_c$ and $J_4$. Actually, the set of pseudo-spin parameters \{$J_{zz},J_\pm,J_{\pm\pm},J_{z\pm}$\} includes the dipolar interaction limited to first neighbours, whereas our calculation makes use of the exchange only integrals and of the infinite range dipolar interaction. Then one must replace ${\cal J}_a$, ${\cal J}_b$ and ${\cal J}_c$ in relations (\ref{snart}) respectively by ${\cal J}_a-D$, ${\cal J}_b-D$ and ${\cal J}_c +2D$ (see Appendix), where $D$ is the characteristic first neighbour dipolar energy worth 0.022\,K in \erti. One obtains, taking into account the error bars given in Ref.\cite{savary} for the pseudo-spin parameters: ${\cal J}_a$= 0.070 $\pm$ 0.048\,K, ${\cal J}_b=-$0.055 $\pm$0.01\,K, ${\cal J}_c=-$0.075 $\pm$0.060\,K and $J_4=-$0.018 $\pm$ 0.015\,K. Except for ${\cal J}_b$, the range of acceptable values is quite large. The isothermal magnetisation being quite sensitive to the particular values of the exchange parameters, the calculation of its field variation at 0.15\,K for {\bf H} // [111], [110] and [100] allows us to precise their values and to reduce significantly the error bars. We obtain the best match to the data (see Fig.\ref{ertimh}) with the following parameter values: ${\cal J}_a$= 0.030 $\pm$0.005\,K, ${\cal J}_b=-$0.050 $\pm$0.005\,K, ${\cal J}_c=-$0.105 $\pm$0.01\,K and $J_4=\pm$0.005\,K. The slight upward curvature at low field for {\bf H} // [111] and [110] and its absence for {\bf H} // [100], the critical field values and the linear increase of the magnetisation above $H_c$ are well reproduced. The derived exchange parameter values, while lying within the range determined in Ref.\cite{savary} from the fit of the spin wave dispersion laws, represent a much more precise set for the exchange tensor. 

We note that, with these exchange integral values, the ground configuration in zero field is found to be the so-called antiferromagnetic $\psi_2$ state, as determined from neutron diffraction \cite{champion,poole}. This result, which is obtained using mean field theory alone including the total crystal field interaction, had been evoked in Ref.\cite{clarty}. However, we find that the Er moments in the $\psi_2$ state do not lie exactly in their easy plane, but have a small ($\simeq$ 0.01\,\mub) out-of-plane $z$-component. The associated mean field N\'eel temperature is $T_{\rm N}$= 2.62\,K, higher than the experimental value 1.2\,K. This enhancement of the mean field $T_{\rm N}$ value with respect to the actual value is attributed in Ref.\cite{savary} to the effect of classical spin fluctuations.

\begin{figure}
\epsfxsize=430pt
\center{
\epsfbox{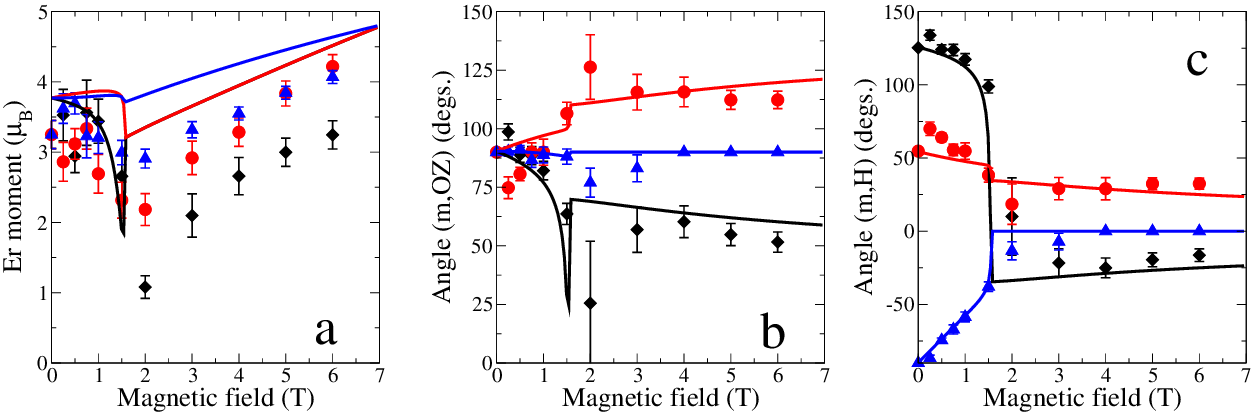}
\caption{\label{erti110} In \erti\ at 0.3\,K, with the magnetic field applied along [110]: {\bf a}) moduli of the Er moments for the 4 sites; {\bf b}) angles of the 4 Er moments with their local ternary axis $<$111$>$; {\bf c}) angles of the 4 Er moments with the field {\bf H}. The data are from Ref.\cite{cao10} and the simulations (solid lines) were performed with the anisotropic exchange tensor ${\cal J}_a =0.030$\,K, ${\cal J}_b=-0.050$\,K, ${\cal J}_c=-0.105$\,K and $J_4$ = 0. The red and black data points and lines correspond to $\alpha$-sites, the blue data points and lines to the two $\beta$ sites, which have the same behaviour as a function of field. Note in {\bf c}) that the moment at the $\alpha$-site with local axis [111] (black line) reverses its direction near the critical field.}
}
\end {figure}
The calculated field variations of the moduli of the 4 Er moments in a tetrahedron are represented in the insets of Fig.\ref{ertimh}. For {\bf H} // [111], the Er moment at the site with its ternary axis parallel to the field has a rather simple behaviour. Its modulus (black curve in the inset of Fig.\ref{ertimh} {\bf a}) is seen to strongly decrease as H increases, signalling a departure from the easy plane; it reaches a minimum value near the critical field $H_c\simeq$1.6\,T, and increases linearly with $H$ above $H_c$. For {\bf H} // [110], a similar behaviour is obtained for the $\alpha$-sites (black and red curve in the inset of Fig.\ref{ertimh} {\bf b}). For both these field directions, there are three different behaviours for the moments as a function of field, which implies that the field direction is not a symmetry axis for the magnetic structure. By contrast, for {\bf H} // [100] (Fig.\ref{ertimh} {\bf c}), a single behaviour is obtained for the 4 sites, suggesting that [100] is a symmetry axis for the field induced structure, which could be checked by in-field neutron diffraction.
\begin{figure}
\center{\epsfxsize=170pt\epsfbox{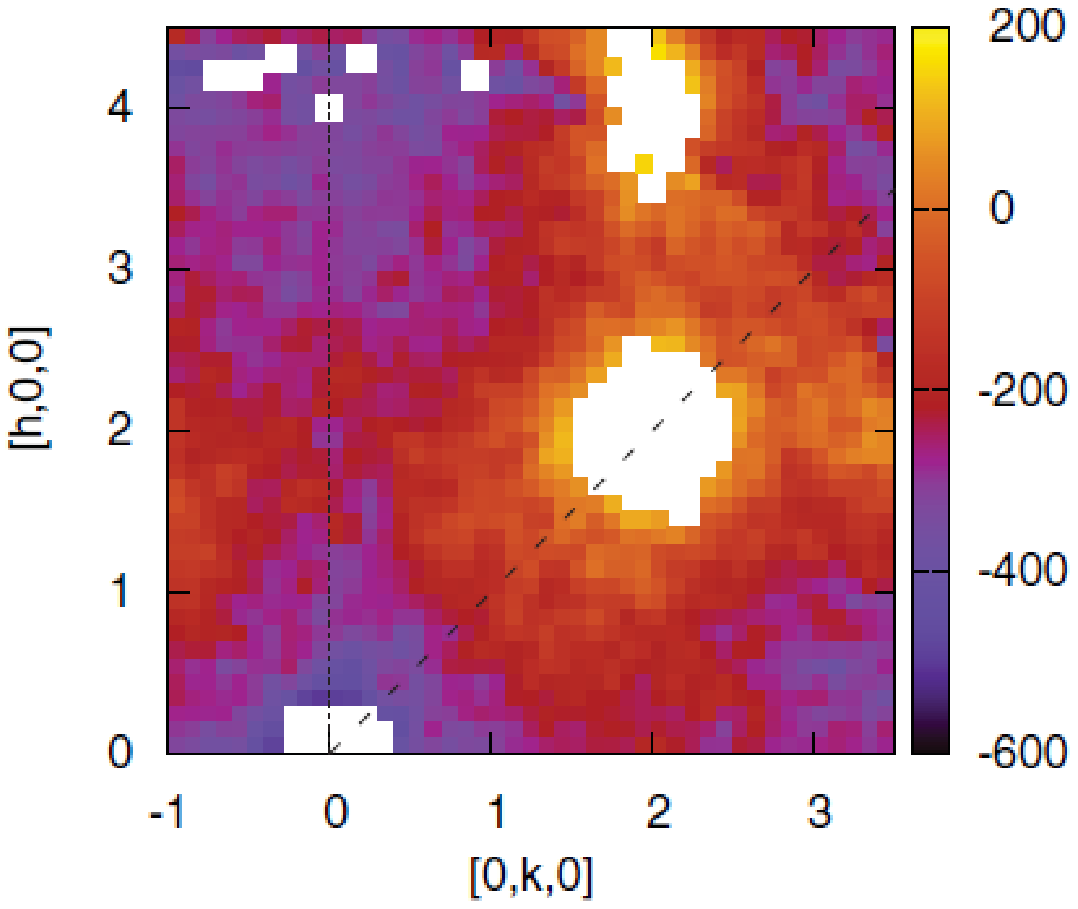}
        \epsfxsize=240pt\epsfbox{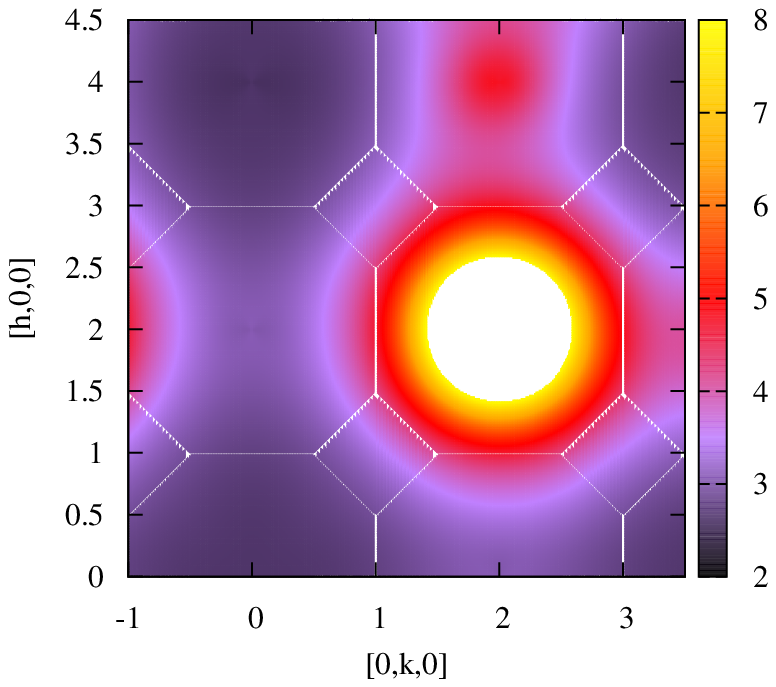}} 
\center{\epsfxsize=200pt\epsfbox{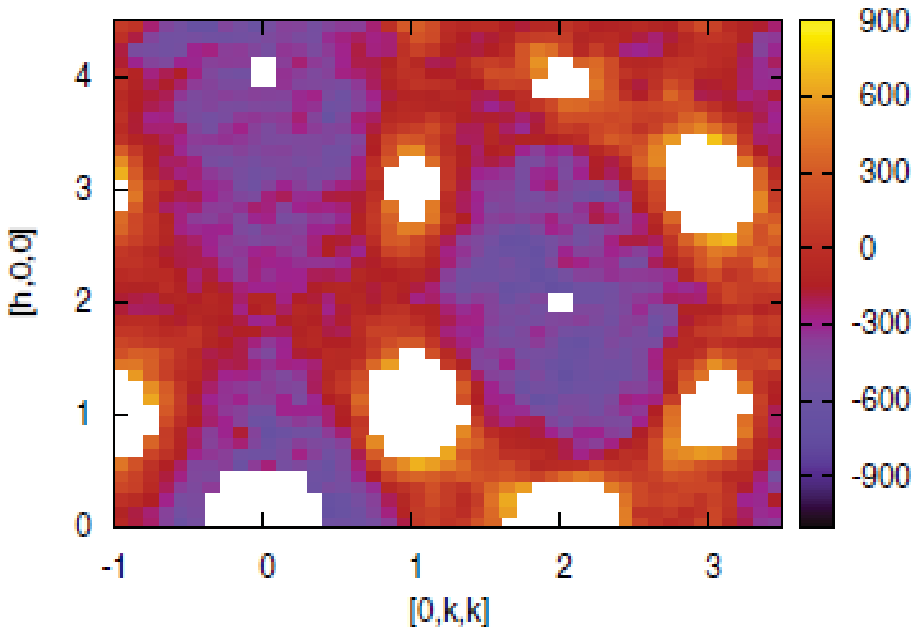}
        \epsfxsize=220pt\epsfbox{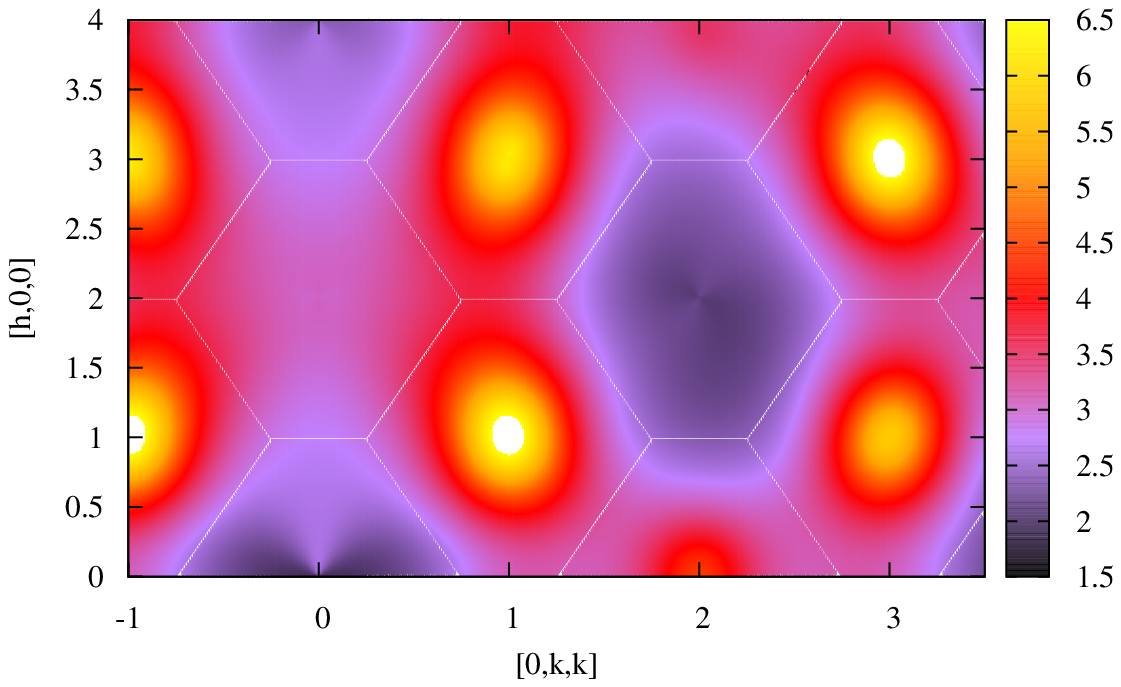}}
\caption{\label{diffneutr} Diffuse neutron scattering at 2\,K in \erti\ in the $(hk0)$ plane (upper panel) and in the $(hkk)$ plane (lower panel) of the reciprocal space: {\bf Left:} experimental data from Ref.\cite{dalm_curnoe}; {\bf Right:} simulated maps (at 3.2\,K) using the exchange tensor ${\cal J}_a=$0.030\,K, ${\cal J}_b=-$0.050\,K and ${\cal J}_c=-$0.105\,K in the 4-site RPA approximation.}
\end{figure}

Neutron diffraction was performed at 0.3\,K for {\bf H} // [110] in \erti\ \cite{ruff,cao10} and the data were interpreted in Ref.\cite{cao10} using, as a crude approximation, a two-component exchange tensor diagonal in the local frame. We replot here the data of Ref.\cite{cao10} (Fig.\ref{erti110}) and show that the evolution of the magnetic structure can be reproduced using the presently derived 3-component exchange tensor diagonal in the bond frames (neglecting the Dzyaloshinski-Moriya exchange term). This gives a physical insight about the field evolution of the magnetic structures, which present similar features whatever the field direction. On Fig.\ref{erti110} {\bf c}, which depicts the field variation of the angles between the Er moments and the field, it is clear that the moments rotate towards the field direction up to the critical field value, then align along (or close to) the field. The critical field can thus be considered as the ``spin flip'' field of the AF structure, and the linear increase of the moment moduli (and of the magnetisation) above $H_c$ is caused by quantum mixing with the excited crystal field states. The calculated values of the angles are in good agreement with the data (Fig.\ref{erti110} {\bf b} and {\bf c}), but the calculated moment moduli above $H_c$ (Fig.\ref{erti110} {\bf a}) are somehow overestimated.

The neutron diffuse scattering in the paramagnetic phase at 2\,K reported in Ref.\cite{dalm_curnoe} is displayed in the left part of Fig.\ref{diffneutr}. It shows intense spots near the Bragg positions (111), (133), (311), (333) ..., and less intense maxima near (000), (022) ... Our calculations (Fig.\ref{diffneutr} right part) of this diffuse scattering are performed using the RPA approximation \cite{kao} with the anisotropic exchange tensor derived above from the magnetisation curve. The temperature was chosen at 3.2\,K since the mean field N\'eel temperature with these parameters is 2.62\,K. One can see that our simulations reproduce satisfactorily the experimental data of Ref.\cite{dalm_curnoe}. We note that they capture the high intensity diffuse spots at Bragg positions, precursor to the long range magnetic ordering with {\bf k} = 0 occuring at lower temperature, whereas the simulations performed in Ref.\cite{dalm_curnoe} do not. We believe this is due to the fact that the approach used in this latter work, i.e. the diagonalisation of the 4-site exchange/dipolar hamiltonian on a tetrahedron, cannot describe the critical correlations above $T_{\rm N}$ responsible for these diffuse Bragg spots.

\section{Very low temperature isothermal magnetisation in \tbti.}

In the spin liquid \tbti, where long range magnetic ordering does not occur down to very low temperature in zero field, application of a magnetic field induces magnetic order with {\bf k}=0 for {\bf H} // [110] \cite{sazonov} and {\bf H} // [111] \cite{sazguk}, and probably also for {\bf H} // [001] although no neutron diffraction data are available for this field direction. Actually, for {\bf H} // [110], an AF structure with {\bf k} = [100] coexists with the {\bf k}=0 structure above 2\,T and below 1\,K \cite{sazonov}, but it is not expected to contribute to the magnetisation. Then, the 4-site self-consistent calculation sketched in the previous section is expected to hold for describing the magnetisation associated with the field induced magnetic structure, but a knowledge of the zero field ground state in \tbti\ is necessary for this purpose. We recently proposed a model which can account for the zero-field spin liquid phase \cite{bonv11,bonville}, and which involves the development, at low temperature, of a symmetry breaking at the rare earth site. This symmetry breaking could be either an effect precursor to a Jahn-Teller transition occuring at much lower temperature or due to quadrupolar ordering, as recently suggested for non-Kramers ions \cite{onoda1}. The associated distortion, assumed to be of tetragonal symmetry \cite{bonville,onoda1}, reads in the local frame, taking for instance its axis along the cubic [001] axis:
\begin{equation}
{\cal H}_Q = \frac{D_Q}{3} \ [2J_x^2+J_z^2+\sqrt{2}\ (J_xJ_z+J_zJ_x)],
\label{dist}
\end{equation}
where $D_Q$ is the strength of the distortion. The main effect of the symmetry breaking is to lift the degeneracy of the ground crystal field doublet and to destroy the Ising character of its wave-functions. It also results in the appearance, in the space of AF exchange integrals \{${\cal J}_a,{\cal J}_b,{\cal J}_c$\}, of regions where the mean field Tb moment is zero near zero temperature, i.e. of a spin liquid phase where short range correlations alone are present. The distortion strength and the exchange tensor derived in Ref.\cite{bonville}, which describe correctly the temperature variation of the local susceptibility \cite{cao09}, the variation of the field-induced magnetic structure \cite{sazonov} and the inelastic and diffuse neutron scattering at very low temperature \cite{petit} in \tbti, are respectively: $D_Q$ = 0.25\,K and ${\cal J}_a=-$0.07\,K, ${\cal J}_b = -$0.19\,K and ${\cal J}_c=-$0.09\,K. In a single crystalline sample, one expects the tetragonal distortion to be distributed in domains with axis along the 3 fourfold cubic axes. Another feature in \tbti\ is the magneto-elastic (ME) interaction which yields giant magnetostriction effects, due to the presence of the low lying crystal field level \cite{aleks,kleko}. Averaging over the domains and taking into account the magneto-elastic interaction (limited to quadratric terms in the total angular momentum) following the formalism of Ref.\cite{kleko}, we have computed the 0.08\,K and 4\,K magnetisation for fields along the 3 symmetry directions [100], [110] and [111]. We used the above quoted anisotropic exchange tensor and the trigonal crystal field as in Ref.\cite{cao09}, in the presence of the symmetry breaking (\ref{dist}). At 4\,K, we set the distortion to zero since, at this temperature, its strength should be much smaller, but probably non-vanishing. 
\begin{figure}
\epsfxsize=325pt
\center{\epsfbox{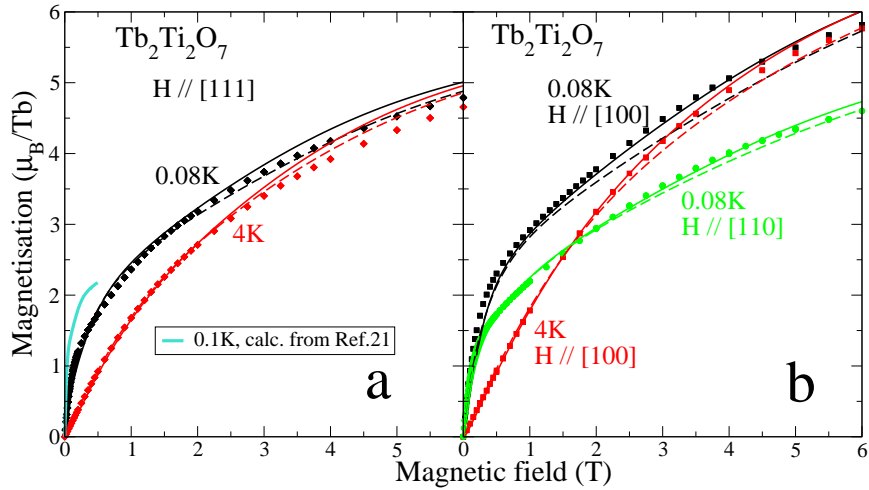}
\caption{\label{mh3axe} Isothermal magnetisation curves in \tbti\ at 0.08\,K and 4\,K: {\bf a} for {\bf H} // [111], {\bf b} for {\bf H} // [100] and [110]. The experimental data for {\bf H} // [111] and [110] are taken from Ref.\cite{lhotel}. The calculated curves were obtained using our model (tetragonal distortion and anisotropic exchange) with two assumptions: no magneto-elastic effects (dashed lines) and including the magneto-elastic distortions (solid lines), except for the solid blue line in {\bf a} which reproduces the prediction of the ``quantum spin ice'' model at 0.1\,K of Ref.\cite{molavianging}, up to the highest field of the calculation (0.5\,T).}
}
\end{figure}

The comparison between experimental data and our calculation is shown in Fig.\ref{mh3axe}. The overall agreement is reasonably good for both temperatures 0.08\,K and 4\,K. In order to assess the importance of the ME interactions, we have represented the calculated curves obtained without (dashed lines) and with (solid lines) ME effects. Inclusion of the ME interaction slightly modifies the magnetisation, especially at high fields. It yields a small enhancement, which results in a better agreement with experiment for {\bf H} // [100] and [110] (Fig.\ref{mh3axe} {\bf b}), but not for {\bf H} // [111] (Fig.\ref{mh3axe} {\bf a}). For this latter field direction, we have also reproduced the curve calculated using the ``quantum spin ice'' model at 0.1\,K (blue line), taken from Ref.\cite{molavianging}. Despite the limited field range, it is clear that it does not reproduce the data, which invalidates the ``quantum spin ice'' model for \tbti, at least as far as the magnetisation is concerned. Furthermore, this model predicts a sizeable variation of the shape of the low field magnetisation curve between 0.02\,K and 0.1\,K, with the appearance of a clear plateau at the lowest temperatures (0.02\,K). By contrast, the experimental data \cite{dunsiger,lhotel} (and our calculations) show that the shape of the magnetisation curve does not appreciably change between 0.05\,K and 0.3\,K. We believe the absence in \tbti\ of the magnetisation plateau expected for Ising spins is due to the proposed symmetry breaking at the rare earth site, which destroys the Ising character linked with the bare trigonal crystal field wave-functions.

Although our model correctly reproduces the overall magnetisation behaviour and its anisotropy, some deviation occurs at 0.08\,K at low field, around 1\,T and below, mainly for {\bf H} // [111] and [100]. The curvature of the magnetisation as the field increases is not exactly reproduced. At higher fields, above 3-4\,T, the calculated points lie somewhat above the data points, especially for {\bf H} // [111] both at 0.08\,K and 4\,K. Keeping in mind that the uncertainty for such magnetic measurements is usually estimated to amount to a few percent, we can envisage various causes for these deviations. First, our model would not capture all the details of the field-induced magnetic structure, especially at low fields. Second, the effect of a slight field misalignment with respect to the crystal axes can also play a role. For {\bf H} // [110], indeed, there occurs a ``spin melting'' near 1\,T where the two moments lying on sites with their ternary axis perpendicular to the field ($\beta$ sites) vanish \cite{sazonov}. The occurence of the ``spin melting'' and the configuration of the $\beta$ moments are very sensitive to the alignment of the field with respect to the crystal axis. For {\bf H} // [111], a ``spin melting'' seems also to occur, but only for one of the 4 Tb sites and at a much lower field \cite{sazguk}, and a small misalignment of the field should have practically no effect on the computed curves. Finally, we made an approximation concerning the tetragonal distortion and the magneto-elastic interaction. We  assumed that they can be treated independently, i.e. that the field induced strains merely superimpose onto the tetragonal field independent strain. This implies that the distortion domains are equiprobable and that they are not affected by the magnetic field. At low field, where the ME term in the hamiltonian has a magnitude of a few 10$^{-2}$\,K, to be compared with $D_Q = 0.25$\,K, this assumption seems valid, but at high field where the ME interaction can reach a few 0.1\,K, it could be questionable. A more elaborate theory would be needed to account for these effects, including eventually higher order terms and/or other interactions like the quadrupole-quadrupole coupling, and their field dependence.   

\section{Conclusion}

We have presented a mean field approach for frustrated pyrochlore systems which takes into account the full crystal field level scheme of the rare earth ion, the anisotropic nearest neighbour exchange and the infinite range dipolar interaction. The model is applied to investigate the high field low temperature magnetisation curves in the antiferromagnet \erti\ and the spin liquid \tbti, together with other physical properties. In \erti, we show the equivalence of the exchange tensor diagonal in the bond frame we use here and of the pseudo-spin S=1/2 exchange tensor recently proposed. Our fit of the magnetisation curve in the AF phase, where the quantum mixing with excited crystal field states is well accounted for, allows us to derive much more precise values for the exchange parameters. These parameters also describe well the diffuse neutron scattering in the paramagnetic phase. In \tbti, we introduced a modified crystal field with a small tetragonal distortion from trigonal symmetry and we took into account the magneto-elastic interaction. We show that the exchange tensor we derived in previous works allows the low temperature magnetisation curves for high symmetry field directions to be reasonably well reproduced without further parameters. The advantage of the present approach with respect to models dealing with S=1/2 pseudo-spins lies in the fact that it can take into account the effects linked with quantum mixing of crystal field states by the magnetic field. This is important in \erti\ when dealing with high field properties, and still more important in \tbti\ since the first crystal field excitation has an energy of $\simeq$15\,K and mixing is important even for moderate magnetic fields. 
 
\ack{We are grateful to P. Dalmas de R\'eotier and C. Marin (CEA Grenoble SPSMS, France) for providing the \erti\ single crystal. One of us (P.B.) gratefully acknowledges the help of B. Z. Malkin for the magneto-elastic calculations.}

\appendix
\section{Relations between the exchange tensor appropriate to the S=1/2 pseudo-spin hamiltonian and the exchange tensor in the ``bond frame''}

We start by choosing a set of vectors forming an orthonormal frame attached to the bond linking rare earth neighbours $i$ and $j$. Calling $\vec{e}_i$ the unit vector along the $<111>$ trigonal axis at site $i$, we define the following unit vectors forming the ``bond frame'', where $\vec c_{ij}$ lies along the link between the two neighboring sites :
\begin{eqnarray}
{\bf c}_{ij} &=& \frac{\sqrt{3}}{2 \sqrt{2}} ({\bf e}_j-{\bf e}_i) \nonumber \\
{\bf a}_{ij} &=& \frac{\sqrt{3}}{2} ({\bf e}_i +{\bf e}_j)  \\
{\bf b}_{ij} &= &{\bf c}_{ij} \times {\bf a}_{ij}. \nonumber
\end{eqnarray}
In the bond frame, we consider the general anisotropic exchange Hamiltonian, written in terms of the full angular momenta ${\bf J}$ and where the last term is the antisymmetric Dzyaloshinski-Moriya exchange (the convention used here is that an antiferromagnetic exchange integral is negative) : 
\begin{eqnarray}
{\cal H} = -[ &\sum_{<ij>} &
{\cal J}_a ({\bf J}_i.{\bf a}_{ij})({\bf a}_{ij}.{\bf J}_{j})+
{\cal J}_b ({\bf J}_i.{\bf b}_{ij})({\bf b}_{ij}.{\bf J}_{j}) \nonumber \\
&+&{\cal J}_c ({\bf J}_i.{\bf c}_{ij})({\bf c}_{ij}.{\bf J}_{j})
+\sqrt{2}\ J_4 \ {\bf b}_{ij}.{\bf J}_i \times {\bf J}_{j}],
\end{eqnarray}
where $\sum_{<ij>}$ means a summation over the first neighbour pairs. This can be written in terms of a global exchange matrix $\tilde {\cal J}$:
\begin{eqnarray} \label{hambond}
{\cal H} & = & -\sum_{<ij>,uv} J_i^u \left( {\cal J}_a\  a^u_{ij}~a^v_{ij}+{\cal J}_b\ b^u_{ij}~ b^v_{ij} + {\cal J}_c\  c^u_{ij}~c^v_{ij} + \sqrt{2} \ J_4 \ \eta_{uv} \right) J_j^v \nonumber \\
& = & -\sum_{<ij>}\ {\bf J}_i\ \tilde {\cal J}\ {\bf J}_j,
\end{eqnarray}
where $\tilde \eta$ is an antisymmetric $3 \times 3$ matrix with only two non-zero elements: $\eta_{13}$=1 and $\eta_{31}=-$1. In the ``bond frame'', the matrix $\tilde {\cal J}$ writes:
\begin{equation}\label{jbond}
{\cal J} = \left(
\begin{array}{ccc}
{\cal J}_a & 0 & \sqrt{2}\ J_4 \\
 0 & {\cal J}_b & 0 \\
 -\sqrt{2}\ J_4 & 0 & {\cal J}_c
\end{array}
\right).
\end{equation}

In Ref.\cite{savary}, the general anisotropic exchange Hamiltonian is written in terms of the spin components {\it within the local frame} of an S=1/2 pseudo-spin:
\begin{eqnarray}\label{hpseud}
{\cal H} = \sum_{i,j} {\sf J}_{zz} {\sf S}^z_i {\sf S}^z_j - {\sf J}_{\pm} \left({\sf S}^+_i {\sf S}^-_j + {\sf S}^-_i {\sf S}^+_j + \right) \nonumber \\
+ {\sf J}_{\pm\pm} \left(\gamma_{ij} {\sf S}^+_i {\sf S}^+_j + \gamma^*_{ij} {\sf S}^-_i {\sf S}^-_j \right) + {\sf J}_{z \pm} \left[ {\sf S}_i^z \left( \zeta_{ij} {\sf S}^+_j + \zeta^*_{ij} {\sf S}^-_j\right) + i \leftrightarrow j \right], 
\end{eqnarray}
where the "sanserif" notations refer to spin components in the local bases. The two states of this S=1/2 pseudo-spin span the states of the ground CEF doublet and, for the case of a Kramers ion (like \er), one can define a local g-tensor $\tilde g$ such that: $g_{\rm J} {\bf J} = \tilde g {\sf {\bf S}}$. In the context of pyrochlores with local trigonal symmetry, the $\tilde g$ matrix is diagonal and takes the form:
\begin{equation}
\tilde g = \left(
\begin{array}{ccc}
g_{\perp} & 0 & 0 \\ 0 & g_{\perp} & 0 \\0 & 0 & g_z
\end{array}
\right).
\end{equation}

Our goal is to determine the relation between the ``bond frame'' exchange parameters ${\cal J}_a$, ${\cal J}_b$, ${\cal J}_c$ and $J_4$, on the one hand, and the pseudo-spin parameters ${\sf J}_{zz}$, ${\sf J}_{z\pm}$, ${\sf J}_\pm$ and ${\sf J}_{\pm\pm}$ on the other hand. To this end, we call $B_{ij}$ (resp. $M_i$) the matrix transforming the coordinates in the ``bond frame'' \{$ij$\} (resp. the local frame $i$) to the cubic (cartesian) coordinates, and $A$ the matrix transforming $({\sf S_x},{\sf S_y},{\sf S_z})$ into $({\sf S_+},{\sf S_-},{\sf S_z})$ in the local basis (we omitted the site indexes for sake of clarity): $g_{\rm J}{\bf J} = \tilde g ~A~{\sf \vec{S}}$, with ${\sf \vec{S}}=({\sf S_+},{\sf S_-},{\sf S_z})$, and:
\begin{equation}
A = \left(
\begin{array}{ccc}
1/2 & 1/2i& 0 \\
-1/2i& 1/2 & 0 \\
0 & 0 & 1
\end{array}
\right).
\end{equation}
A straightforward transformation of hamiltonian (\ref{hambond}) shows that:
\begin{equation}
{\cal H} = \sum_{<ij>,uv} {\sf S}_i^u~\left( \frac{1}{g_{\rm J}^2} \ A^T ~\tilde g~M^T_i ~B_{ij}~\tilde {\cal J}~ B_{ij}^T~M_j~ \tilde g ~A \right)^{uv}~{\sf S}_j^v,
\end{equation}
This allows the relations between the two sets of parameters appearing in (\ref{jbond}) and (\ref{hpseud}) to be determined; they are given in section \ref{anis} of the main text. The conventions as to the sign of the exchange integrals are different for the two sets: for an AF interaction, they  are positive (except $J_\pm$) in the pseudo-spin representation and negative in the ``bond frame'' description.

We note that, in the ``bond frame'', the dipolar interaction limited to a nearest 
neighbor pair \{$ij$\}: ${\cal H}_{dip}^{ij} = D \left[ {\bf J}_i {\bf J}_j - 3\  ({\bf J}_i.{\bf c}_{ij})~ ({\bf c}_{ij}.{\bf J}_j) \right]\ = {\bf J}_i \ \tilde D\  {\bf J}_j$ is diagonal with: 
\begin{equation}
\tilde D = \left(
\begin{array}{ccc}
D & 0 & 0 \\
0 & D & 0 \\
0 & 0 & - 2D
\end{array}
\right),
\end{equation}
and $D = \frac{\mu_0}{4\pi}\ \frac{(g_{\rm J} \mu_B)^2}{k_{\rm B}}\ \frac{16\sqrt{2}}{a^3}$,
where $a$ is the cubic cell lattice parameter.

\end{document}